\documentclass[a4paper,11pt]{article}

\usepackage{amsmath,amssymb} 
\usepackage{graphicx,psfrag,subfigure}

\def\ud{\text{d}}

\def\ue{\text{e}}

\title{\LARGE Scalar Perturbation of Black Holes\\ in (2+1)-Dimensional Dilaton Gravity}

\author{Biplab Raychaudhuri\thanks{Department of Physics, Shiksha Bhavan, Visva-Bharati University, Santiniketan, India. E-mail: biplabphy@visva-bharati.ac.in} and Kshounish Guha\thanks{Banizerhat High School, Satkhamar, Jalpaiguri, West Bengal, India. E-mail: kshounish.guha@gmail.com}}

\begin{document}

\date{}

\maketitle

%%%%%%%%%%%%%%%%%%%%%%%%%%%%%%%%%%%%%%%%%%%%%%%%%%%%%%%%%%%%%%%%%%%%%%

\begin{abstract}

%Perturbation of Black Holes in (2+1)-Dimensional Dilaton Gravity.
%Corrected and edited freshly (23.10.10), editing. by BR on 06-- nov
%Checked and Edited by KG on 15 jan
%final form (expected) by BR on 21 jan

Scalar perturbation represented by the Klein-Gordon field equation of the system of black holes in 2+1 dimensional Static charged dilaton gravity proposed by Chan and Mann is investigated analytically. The radial equation is analyzed near horizon and at the asymptotic region to show the absence of superradiance instability.
\end{abstract}

%%%%%%%%%%%%%%%%%%%%%%%%%%%%%%%%%%%%%%%%%%%%%%%%%%%%%%%%%%%%%%%%%%%%%%

\section{Introduction}\label{Intro}

The study of lower dimensional gravitational theories has drawn much interest in recent past in theoretical physics. This is primarily because of the potential insight it offers into different areas of general relativity. In the pursuit of lower dimensional theories of gravity, the natural starting point is $(2+1)$-dimensional general relativity. Consequently in recent years general relativity in $(2+1)$ spacetime dimensions has been a great source of fascination for the theorists. The idea that $(2+1)$-dimensional gravity might be a useful arena for investigating more general questions dates back to at least 1963~\cite{starus}, when Staruszkiewicz (and also  in 1966 by Leutwyler~\cite{Leutwyler}) has given a geometrical description of a point particle in (2+1)-dimensional black hole~\cite{carlipbook}. But the power of the $(2+1)$-dimensional model only became clear with the work of Deser, Jackiw, and 't Hooft~\cite{Deser1, Deser2, Hooft, Deser3}. The study of the Einstein theory of gravity in $(2+1)$ spacetime dimensions has been widely recognized as the useful theoretical laboratory to explore the foundations of classical and quantum gravity. It provides an opportunity to model classical and quantum dynamics of black holes with a simpler set of field equations. The usefulness of $(2+1)$-dimensional gravity comes from the fact that it eliminates the technical problems while preserving the conceptual foundations.

The most direct way of approaching lower-dimensional gravity is to begin with a consideration of Einstein's equations in $(D+1)$-dimensions,
\begin{equation}
G_{\mu \nu} = R_{\mu \nu} - \frac{1}{2} g_{\mu \nu} R = 8 \pi G_{D+1} T_{\mu \nu} ,
\end{equation}
where $R_{\mu \nu}$ is the Ricci tensor, $R$ is the Ricci scalar or Curvature scalar, $G_{\mu \nu}$ is the Einstein tensor, $T_{\mu \nu}$ is the Stress-energy tensor and $G_{D+1}$ is the Newton's constant in $(D+1)$-dimensions. For $D\geq3$, a vanishing stress-energy tensor implies a vanishing Einstein tensor $G_{\mu \nu}$; but it does not mean a vanishing Riemann tensor $R_{\mu \nu \alpha \beta}$. It is possible to have a non-zero curvature in regions of spacetime where there is no stress-energy. This feature is lost when $D\leq2$. In $(2+1)$-dimensions the number of independent components of the Riemann curvature tensor $R_{\mu \nu \alpha \beta}$ and the Einstein tensor $G_{\mu \nu}$ are the same (\textit{i.e.}, six). So it is possible to express the Riemann tensor $R_{\mu \nu \alpha \beta}$ completely in terms of the Einstein tensor $G_{\mu \nu}$:
\begin{equation}
R_{\mu \nu \alpha \beta} \equiv \ \epsilon_{\mu \nu \gamma} \ \epsilon_{\alpha \beta  \sigma} \ G^{\gamma \sigma}.
\end{equation}
Vanishing of the latter implies  vanishing of the former. Consequently, the spacetime described by the vacuum solutions to the Einstein equations in $(2+1)$-dimensions are locally flat and there are no gravitational
waves and no interaction between masses. Unlike the case in $(3+1)$-dimensions, in the Einstein theory of gravitation in $(2+1)$-dimensions there is no Newtonian limit to the field equations~\cite{Barrow} and no propagating degrees of freedom. This means that the gravitational influence of the matter sources will be topological in character. For example, a point source will produce a conical structure to the space-time. In $(2+1)$-dimensions the Newtonian potential is always absent and gravitation has no `action at a distance.' The Einstein gravitational constant remains undetermined in (2+1)-dimensions. $(2+1)$-dimensional gravity has been studied as an exactly soluble system at the classical and quantum levels~\cite{Witten1, Witten2, Carlip1}. An intensive discussion about the intriguing properties of $(2+1)$-dimensional general relativity can be found in~\cite{Verbin}.

The end point of gravitationally collapsing matter is a subject of long-standing interest in general relativity. The black hole is one of the most intriguing structures that have ever emerged out of the theory of gravitation. It is a well-known phenomenon in $(3+1)$-dimensions that under certain circumstances gravitational forces can overwhelm all other forces. It causes complete gravitational collapse of a given system of matter. The final destination is the formation of a black hole. The astrophysical implications of this process continue to be an ongoing area of research. Black holes are commonly conceived as places where the effects of gravity are large, surrounded by a region where these effects are asymptotically negligible. 
With its event horizon and/or ergosphere, area theorem, ``theormodynamics'', information loss, mass accretation -- all of which are geometric in origin -- black holes have become object of intense discussion for the physicists during last fifty years (not to say about the popular science and fiction writers). These and other fascinating features of the black hole in $(3+1)$-dimensions made it long desirable to have a black hole in lower dimensions, which could manifest its key features without unnecessary complexity. In $(2+1)$-dimensions, the pure, sourceless gravitational field has no local degrees of freedom, because in three dimensions the Riemann tensor is given algebraically  by the Einstein tensor, which in turn is algebraically determined by the Einstein field equations. The curvature tensor in $(2+1)$-dimensions depends linearly on the Ricci tensor~\cite{carlipbook}. If there is no matter source and no cosmological constant, the Riemann tensor vanishes and space-time is flat. If there is no matter but a cosmological constant $\Lambda$, the Riemann tensor is that of a space of constant curvature $\Lambda/3$. Thus gravity does not vary from place to place and it does not have any wave degrees of freedom~\cite{Brill1}. $(2+1)$-dimensional gravity has no Newtonian limit~\cite{Barrow}. These were some of the reasons why  one might fear that no such interpretation exists for the $(2+1)$-dimensional black hole~\cite{Giddings}. But it came as a surprise when Ba\~nados, Teitelboim and Zanelli~\cite{Banados1} showed that $(2+1)$-dimensional Einstein theory with a negative cosmological constant admits a solution with almost all the usual features of a black hole. It was shown shortly after the discovery of the BTZ solution that this black hole arises naturally from collapsing matter~\cite{MannRoss}.

Ba\~nados, Teitelboim and Zanelli~\cite{Banados1} realized that black holes can exist in standard $(2+1)$-dimensional Einstein-Maxwell theory if a negative cosmological term of anti-de Sitter (AdS) variety $\Lambda = -\frac{1}{l^2}$ is added to Einstein's equations. This added feature allows analytic treatment of some collapse situations. The Ba\~nados-Teitelboim-Zanelli (BTZ) black hole metric~\cite{Banados1} (in `Schwarzschild' coordinates) given by
\begin{equation}\label{BTZmetric}
\ud s^2 = -N^2 \ud t^2 + N^{-2} \ud r^2 + r^2\left(N^\phi \ud t + \ud \phi \right)^2.
\end{equation}
Here $N^2(r)$ and $N^\phi$ are respectively the lapse and shift functions given by
\begin{equation}\label{LapseFn}
N^2(r) = -M + \frac{r^2}{l^2} + \frac{J^2}{4r^2},
\end{equation}
\begin{equation}\label{ShiftFn}
N^\phi = -\frac{J}{2r^2},
\end{equation}
with $  \infty < t  < \infty$, $0  < r < \infty$ and $0 \leq \phi \leq 2\pi$. The BTZ black hole is characterized by the two constants $M$ and $J$ in Eqs.~\eqref{LapseFn}, ~\eqref{ShiftFn}. These two constants are the conserved charges associated with asymptotic invariance under time displacements (mass) and rotational invariance (angular momentum) respectively. These charges are given by the flux integrals through a large circle of spacelike infinity. These two parameters determine the asymptotic behaviour of the solution. In fact, they are the standard ADM mass and angular momentum. The BTZ black hole differs from the Schwarzschild and Kerr solutions in some important respects. Ba\~nados \textit{et. al.}~\cite{Banados2} showed that the metric~\eqref{BTZmetric} is asymptotically anti-de Sitter. The metric has no curvature singularity at the origin. It is stationary and axially symmetric. But the BTZ solution shares many of the essential features of a realistic $(3+1)$-dimensional black hole, including an event horizon, an inner horizon (in the rotating case), and thermodynamic properties. Carlip~\cite{carlipbook,Carlip2} has extensively reviewed the classical and quantum properties of BTZ black hole.

The extensions of BTZ black hole with charge and dilaton has lead to many interesting results. Recent research has indicated that there are a wide class of $(2+1)$-dimensional black holes. These black holes arise as exact solutions to Einstein-Maxwell dilaton theory in $(2+1)$-dimensions. Chan and Mann~\cite{ChanKC} obtained an interesting class of black hole solution in $(2+1)$-dimensional general relativity. The solution represents a one-parameter $(2 > N > 0, N \neq \frac{2}{3})$  family of static charged black holes coupled to a dilaton field $\phi \propto \ln\frac{r}{\beta}$ with a potential term $e^{b\phi}$. The motivation was the extensions of BTZ black hole with charges and dilaton fields. This black hole can be compared to charged dilaton black holes in $(3+1)$-dimensions constructed by Gibbons and Madea~\cite{Gibbons} and Garfinkle, Horowitz and Strominger~\cite{Garfinkle}. Chan and Mann considered the variation of the Einstein-Maxwell-dilaton action  with respect to the BTZ metric, Maxwell and dilaton fields. The resulting metric obtained by them is given by,

\begin{equation}\label{Chanmetric1}
\begin{split}
\ud s^2 = &-\left(-\frac{2M}{N} r^{\frac{2}{N}-1} + \frac{8\Lambda r^2}{(3N-2)N}
 + \frac{8Q^2}{(2-N)N} \right)\ud t^2 \\
    &+ \frac{4r^{\frac{4}{N}-2}}{N^2 \gamma ^\frac{4}{N}} \frac{\ud r^2}{\left(-\frac{2M}{N} r^{\frac{2}{N}-1} + \frac{8\Lambda r^2}{(3N-2)N}
 + \frac{8Q^2}{(2-N)N} \right)} + r^2 \ud \theta ^2.
\end{split}
\end{equation}
Here $\gamma$ is an intregation constant of dimension $[L]^\frac{2-N}{2}$. The metric~\eqref{Chanmetric1} represents the one-parameter family of static charged dilaton black hole in $(2+1)$-dimensions. This black hole solution depends on the parameter $N$ $(2 > N > 0, N \neq \frac{2}{3})$. Chan and Mann pointed that for this family of black holes, the event horizon exists only in the situation $2 > N > \frac{2}{3}$. Within this range there are five possible discrete values of $N = \frac{6}{7}$, $\frac{6}{5}$, $\frac{4}{3}$, $\frac{4}{5}$ and $1$. In the other region where $\frac{2}{3} > N > 0$, the solution has comsological horizon.

In 1996, Chan and Mann came up with another new class of spinning black solutions of $(2+1)$-dimensional gravity minimally coupled to a dilatonic potential of \textbf{$e^{b\phi} \Lambda$}. This class of solutions are asymptotically non-flat. Spinning BTZ metric comes out as a special case of it~\cite{chanmann96}. Chan in 1997 found another class of Black hole solutions in dilatonic and scalar-tensor alternative theories of gravity~\cite{chan97}.

The structure of this paper is as follows. In Sec.~\ref{Intro} we discuss black holes in lower dimensions and introduce the BTZ black hole~\cite{Banados1} and the Chan and Mann static charged dilaton black hole~\cite{ChanKC} in $(2+1)$-dimensions. In Sec.~\ref{Perturbation} we mention briefly the perturbation of black holes and quasinormal modes. Sec.~\ref{ChanPerturbation} focuses upon the perturbation of the static charged dilaton black hole~\cite{ChanKC} in (2+1)-dimensions. In order to study the external scalar wave perturbation of this special class of black hole, we analyze the Klein-Gordon equation in the background of static charged dilaton black hole in $(2+1)$-dimensions and separate the angular and the radial parts. We have made our study general enough by not assigning any specific value to the parameters. In Sec.~\ref{Radial} the radial equation is analysed at asymptotia and near horizon. Finally, we look into the possibility of observing the phenomenon of superradiance in the static charged dilaton black hole~\cite{ChanKC} in $(2+1)$-dimensions.

%%%%%%%%%%%%%%%%%%%%%%%%%%%%%%%%%%%%%%%%%%%%%%%%%%%%%%%%%%%%%%%%%%%%%%
%%%%%%%%%%%%%%%%%%%%%%%%%%%%%%%%%%%%%%%%%%%%%%%%%%%%%%%%%%%%%%%%%%%%%%

\section{Perturbation of Black Holes} \label{Perturbation}
When the evolution of some conservative system is described, one often considers a small departure from a known solution of the system, and generally arrives at a wave equation describing it. For a system with no explicit time dependence, the normal modes are found out as the solutions of the wave equation, satisfying certain boundary conditions. The perturbation can be completely specified as a linear superposition of the normal modes. The operator associated to the perturbation is self-adjoint. So the normal modes are complete and the normal mode frequencies are real. When one deals with open dissipative systems, the expansion of the perturbation in the form of superposition is not possible. In this case instead of normal modes, quasinormal modes (QNM) are to be considered. The frequencies for the quasinormal modes no longer remain purely real, instead they become complex. This indicates that the system loses energy. A precise mathematical definition for a QNM can be given as a pole in the Green's function~\cite{Kokkotas}. But from a more phenomenological point of view, QNM describes a decay of the field under consideration. In general, QNMs are not complete and therefore insufficient to fully describe the dynamics of the system. Nevertheless, QNMs dominate the intermediate stages of the perturbation. Quasinormal modes are therefore extremely important.

Perturbation of black holes has been a topic of much discussion in the last few decades. When a black hole is perturbed, it is found that during a certain time interval the evolution of the initial perturbation is dominated by damped single-frequency oscillation. This kind of perturbation which is damped quite rapidly and exists only in a limited time interval corresponds to quasinormal modes (QNM). Studies of QNMs of perturbations by different fields, such as scalar, electromagnetic, fermionic and gravitational fields, have taken important place in black hole physics. QNMs dominate most of the processes involving perturbed black holes. They give information on the stability properties of black holes. Quasinormal modes of a classical perturbation of black hole spacetime are defined as the solutions to the related wave equations characterized by purely incoming waves at the horizon. In addition, one has to impose boundary condition on the solutions at the asymptotic region. In asymptotically flat spacetime, another boundary condition is necessary for the solution to be purely outgoing at spatial infinity~\cite{Fernando}. QNM frequencies depend only on the black hole properties, such as the mass, angular momentum and charge, and not on the initial perturbations. So QNM frequencies allow a direct way of identifying the spacetime parameters. Quasinormal modes carry a unique fingerprint, which can lead to the direct identification of the existence of the black hole. Detection of these quasinormal modes is expected to be realized through gravitational wave observations. If the radiations due to the QNMs are detected by future gravitational wave detectors, the possible charges of a black hole can be clearly identified. In order to extract as much information as possible from gravitational wave signal, it is important that we understand exactly how the quasinormal modes behave for the parameters of black holes in different models. It must be noted that time-dependent black hole spacetimes can describe the black hole absorption and evaporation processes. Colliding black holes are extremely interesting in this regard. They produce stronger gravitational wave signals which may be easier to detect. Chandrasekhar, in his phenomenal work, \textit{The mathematical Theory of Black holes}, discussed perturbations of black holes in characteristic details~\cite{chandra}. A more recent review of perturbation of black holes can be found in~\cite{Kokkotas} and ~\cite{Nollert}.

%%%%%%%%%%%%%%%%%%%%%%%%%%%%%%%%%%%%%%%%%%%%%%%%%%%%%%%%%%%%%%%%%%%%%%
%%%%%%%%%%%%%%%%%%%%%%%%%%%%%%%%%%%%%%%%%%%%%%%%%%%%%%%%%%%%%%%%%%%%%%

\section{Scalar Perturbation of Static Charged Dilaton Black Hole in $(2+1)$-Dimensions}\label{ChanPerturbation}
In this paper we focus ourselves to the study of the perturbation of the Chan-Mann class of static charged dilaton black hole solutions~\cite{ChanKC}. In this section we shall consider the perturbation of the metric~\eqref{Chanmetric1} by a massive scalar field $\chi$. 

The dynamics of a massive scalar field $\chi$ of mass $\mu$ in the background of a black hole is described by the Klein-Gordon equation,
\begin{equation}
\Box \chi - \mu ^2 \chi = 0,
\end{equation}
which becomes, when written explicitly
\begin{equation}\label{KGEqn1}
\frac{1}{\sqrt{-g}} \partial_\mu \left(\sqrt{-g} g^{\mu\nu} \partial_\nu \chi\right) - \mu ^2 \chi = 0.
\end{equation}

We analyze the quantum mechanical Klein-Gordon field  equation~\eqref{KGEqn1} in the background curved spacetime of static charged dilaton black hole in (2+1)-dimensions using the method of separation of variables. Our study is made general enough by not assigning any specific value to the parameters. Brill \textit{et. al.}~\cite{Brill2} were the first to use the method of separation of variables to find solutions of the scalar wave equation in Kerr black hole background.

The metric for static charged dilaton black hole in (2+1)-dimensions as given in Eq.~\eqref{Chanmetric1} can be expressed as
\begin{equation} \label{Chanmetric2}
\ud s^2 = - \frac{\Delta}{r^2} \ud t^2 + \frac{r^2 \alpha(r)}{\Delta} \ud r^2 + r^2 \ud \theta^2.
\end{equation}
Here \begin{equation}\label{Delta}
\Delta = \left[-\frac{2M}{N} r^{\frac{2}{N}+1} + \frac{8 \Lambda r^4}{(3N-2)N} + \frac{8 Q^2 r^2}{(2-N)N}\right],
\end{equation}
and \begin{equation}\label{alpha(r)}
\alpha(r) = \frac{4r^{\frac{4}{N}-2}}{N^2 \gamma^\frac{4}{N}}.
\end{equation} 
The Klein-Gordon equation~\eqref{KGEqn1} now leads to
\begin{equation} \label{KGEqn2}
-\frac{r^4}{\Delta} \frac{\partial^2\chi}{\partial t^2} + \frac{r^2}{r \sqrt{\alpha(r)}} \frac{\partial}{\partial r} \left(\frac{\Delta}{r \sqrt{\alpha(r)}} \frac{\partial \chi}{\partial r}\right) + \frac{\partial^2 \chi}{\partial \theta ^2} - \mu ^2 r^2 \chi = 0.
\end{equation}
In order to solve equation~\eqref{KGEqn2} we use the separation ansatz of the form
\begin{equation} \label{Ansatz}
\chi\left(r, \theta, t\right) = \frac{1}{r^{1/2}} R\left(r\right) \Theta\left(\theta\right) e^{-iwt}.
\end{equation}
Substituting the ansatz~\eqref{Ansatz} in Eq.~\eqref{KGEqn2} and dividing both sides by $\chi\left(t, r, \theta \right)$ we obtain
\begin{equation}
\frac{w^2 r^4}{\Delta} + \left(\frac{r^{1/2}}{R}\right) \frac{r^2}{r \sqrt{\alpha(r)}}\frac{\partial}{\partial r} \left(\frac{\Delta}{r \sqrt{\alpha(r)}} \frac{\partial}{\partial r} \left(\frac{R}{r^{1/2}}\right)\right) - \mu ^2 r^2 + \frac{1}{\Theta}\frac{\partial^2 \Theta}{\partial \theta ^2} = 0.
\end{equation}
We have grouped the different terms to show explicitly the separability of this equation. With the use of separation constant $\lambda$, we find the separated homogeneous equations in radial and angular coordinates :
\begin{equation} \label{KGEqn3}
\frac{w^2 r^4}{\Delta} + \left(\frac{r^{1/2}}{R}\right) \frac{r^2}{r \sqrt{\alpha(r)}}\frac{\partial}{\partial r} \left(\frac{\Delta}{r \sqrt{\alpha(r)}} \frac{\partial}{\partial r} \left(\frac{R}{r^{1/2}}\right)\right) - \mu ^2 r^2 = - \frac{1}{\Theta}\frac{\partial^2 \Theta}{\partial \theta ^2} = \lambda.
\end{equation}

We will first consider the angular equation. In the next subsection we move our attention to the radial equation.

%%%%%%%%%%%%%%%%%%%%%%%%%%%%%%%%%%%%%%%%%%%%%%%%%%%%%%%%%%%%%%%%%%%%%%%%%%%%%%%%%%%%%%%%%%%%%%%%%%%%%

\subsection{Angular Equation}

The angular part obtained by separating Eq.~\eqref{KGEqn3} finally takes the following form,
\begin{equation} \label{Angular}
\frac{\ud^2 \Theta}{\ud \theta ^2} + \lambda \Theta = 0.
\end{equation}

This equation represents the well-known wave equation in $\Theta$. Here it must be noted that some authors have considered the wave form of $\theta$ in separating the scalar wave equation~\cite{Fernando}. But  we have taken separation ansatz in the form as in Eq.~\eqref{Ansatz} for the sake of generality. The final form of the angular equation~\eqref{Angular} we obtained confirms that the angular variable will have a wave like behaviour.

%%%%%%%%%%%%%%%%%%%%%%%%%%%%%%%%%%%%%%%%%%%%%%%%%%%%%%%%%%%%%%%%%%%%%%%%%%%%%%%%%%%%%%%%%%%%%%%%%%%%%

\subsection{Radial Equation}
The radial part obtained by separating Eq.~\eqref{KGEqn3} is
\begin{equation}
\frac{w^2 r^4}{\Delta} + \left(\frac{r^{1/2}}{R}\right) \frac{r^2}{r \sqrt{\alpha(r)}}\frac{\ud}{\ud r} \left(\frac{\Delta}{r \sqrt{\alpha(r)}} \frac{\ud}{\ud r} \left(\frac{R}{r^{1/2}}\right)\right) - \mu ^2 r^2 - \lambda = 0.
\end{equation}
After multiplying by $\frac{\Delta}{r^2}$ and rearranging suitably the above equation takes the form
\begin{equation} \label{RadialEqn1}
\left(\frac{r^{1/2}}{R}\right) \frac{\Delta}{r \sqrt{\alpha(r)}}\frac{\ud}{\ud r} \left(\frac{\Delta}{r \sqrt{\alpha(r)}} \frac{\ud}{\ud r} \left(\frac{R}{r^{1/2}}\right)\right) + w^2 r^2 - \mu ^2 \Delta -  \frac{\lambda \Delta}{r^2} = 0.
\end{equation}
For the analysis of the radial equation~\eqref{RadialEqn1}, we introduce the so-called tortoise coordinate $r^*$ defined as
\begin{equation} \label{Tortoise} 
\frac{\ud r^*}{\ud r} = \frac{r^2 \sqrt{\alpha(r)}}{\Delta}.
\end{equation}
We thus obtain
\begin{equation}
\frac{\Delta}{r \sqrt{\alpha(r)}} \frac{\ud }{\ud r} = r \frac{\ud }{\ud r^*}.
\end{equation}

The tortoise coordinate $r^*$ spans the entire real line while the usual radial coordinate $r$ spans only the half-line. The use of tortoise coordinate pushes the horizon off to minus infinity, \textit{i.e.} the horizon $\Delta \rightarrow 0$ maps to $r^* \rightarrow -\infty$. On the other hand, $r \rightarrow \infty$ corresponds to $r^* \rightarrow +\infty$.

The radial equation~\eqref{RadialEqn1} now takes the form of a second order differential equation in $r^*$ coordinate,
\begin{equation} \label{RadialEqn2}
\begin{split}
\frac{\ud^2 R}{\ud {r^*}^2} & + \left[ \omega^2 - {\frac{\mu^2 \Delta}{r^2}} - {\frac{\lambda \Delta}{r^4}} \right] R + \left[ {\frac{5 \Delta^2}{4 \alpha r^6}} - {\frac{16 \Lambda \Delta}{(3N-2)N \alpha  r^2}} - {\frac{8Q^2 \Delta}{(2-N)N \alpha r^4}} \right] R \\ 
& + \left[ { {\frac{M}{N \alpha}} \left(\frac{2}{N}+1 \right) \frac{\Delta}{r^{(5 - \frac{2}{N})}} } - { \frac{N}{4} \left(1 - \frac{N}{2}\right) \frac{\gamma^{\frac{2}{N}}}{\sqrt \alpha} \frac{\Delta^2}{r^{(\frac{2}{N}+5)}} } \right] R  = 0.
\end{split}
\end{equation}
After proper rearrangement the above equation can be expressed in the form of a one-dimensional Schr\"odinger-type equation in $r^*$ coordinate with a potential function $V(r)$,
\begin{equation}\label{second order}
\frac{\ud^2 R}{\ud {r^*}^2} + \left[ \omega^2 - V(r) \right] R  = 0.
\end{equation}
The Potential function $V(r)$ is given by
\begin{equation}\label{V_r1}
\begin{split}
V(r) = {\frac{\mu^2 \Delta}{r^2}} + & {\frac{\lambda \Delta}{r^4}} - {\frac{5 \Delta^2}{4 \alpha r^6}} + {\frac{16 \Lambda \Delta}{(3N-2)N \alpha  r^2}} + {\frac{8Q^2 \Delta}{(2-N)N \alpha r^4}} \\ 
 - & { {\frac{M}{N \alpha}} \left(\frac{2}{N}+1 \right) \frac{\Delta}{r^{(5 - \frac{2}{N})}} } - { \frac{N}{4} \left( \frac{N}{2} - 1 \right) \frac{\gamma^{\frac{2}{N}}}{\sqrt \alpha} \frac{\Delta^2}{r^{(\frac{2}{N}+5)}} }.
\end{split}
\end{equation}
Now we substitute the value of $\alpha$ in the above equation from equation~\eqref{alpha(r)} to obtain the final form of the Potential function $V(r)$,
\begin{equation}\label{V_r2}
\begin{split}
V(r) = {\frac{\mu^2 \Delta}{r^2}} + & {\frac{\lambda \Delta}{r^4}} - { \frac{5 N^2 \gamma^{\frac{4}{N}}}{16} }{ \frac{\Delta^2}{r^{(\frac{4}{N}+4)}} } + { \frac{4N \Lambda \gamma^{\frac{4}{N}}}{(3N-2)} }{ \frac{\Delta}{r^{\frac{4}{N}}} } + { \frac{2NQ^2 \gamma^{\frac{4}{N}}}{(2-N)} }{ \frac{\Delta}{r^{(\frac{4}{N}+2)}} } \\ 
 - & { {\frac{M (N+2) \gamma^{\frac{4}{N}}}{4}}  \frac{\Delta}{r^{( \frac{2}{N} + 3)}} } - { \frac{N (N - 2) \gamma^{\frac{4}{N}}}{8} \frac{\Delta^2}{r^{(\frac{4}{N}+4)}} }.
\end{split}
\end{equation}

Chan and Mann~\cite{ChanKC} indicated that for black hole solution the parameter $N$ lies in the range $(2 > N > \frac{2}{3})$. For different values of $N$ different situations arise and the potential function $V(r)$ takes different forms.  It is instructiove to draw some  \textit{representative} graphs of the potential function $V(r)$ for some values of $N$ in the range $(2 > N > \frac{2}{3})$. For simplicity we plot potentials where the mass of the scalar field is zero. The other parameters are chosen as follows: $\gamma=1$, $\lambda=1$ and $\Lambda=1$.

%%%%%%%%%%%%%%%%%%%%%%%%%%%%%%%%%
\begin{figure}[htb]
\psfrag{M5Q05}{}
\centering
\subfigure[$M=5, Q= 0.5$]{
\includegraphics[width=0.45\textwidth,height=5cm]{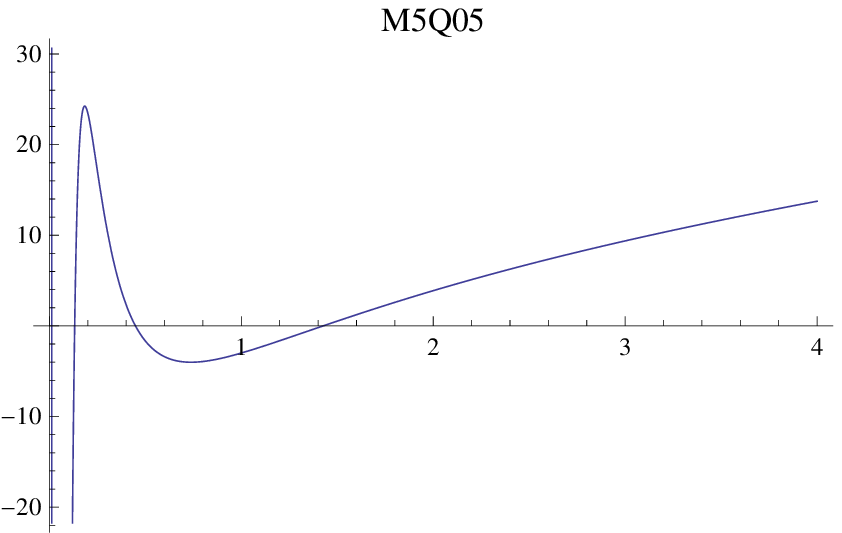}}
\hspace*{\fill}
\subfigure[$M=5, Q=1$]{
\psfrag{M5Q1}{}
%\centering
\includegraphics[width=0.45\textwidth,height=5cm]{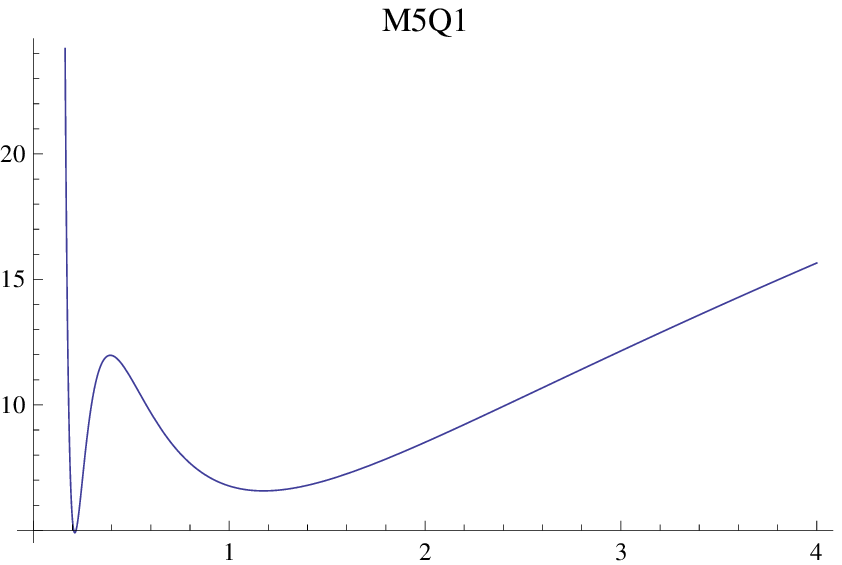}}
\caption{$V(r)$ for $N=\frac{6}{5}$}
\end{figure}

%%%%%%%%%%%%%%%%%%%%%%%%%%%%%%%%%%%%%%%

\begin{figure}[htb]
\centering
%\psfrag{m1q005}{$M=1, Q= 0.05$}
\psfrag{m1q005}{}
\subfigure[$M=1, Q=0.05$]{
\includegraphics[width=0.45\textwidth,height=5cm]{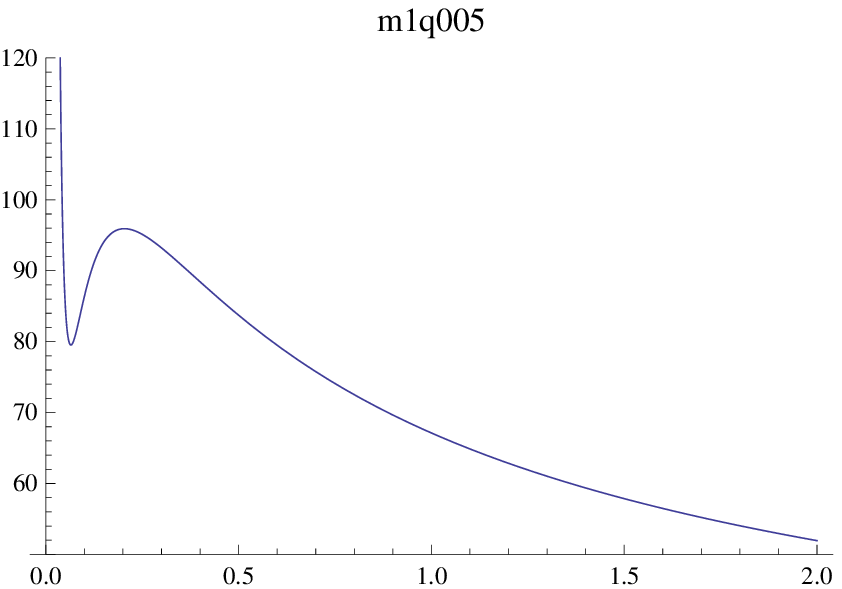}}
\hspace*{\fill}
\subfigure[$M=2, Q=0.1$]{
%\psfrag{m2q01}{$M=2, Q=0.1$}
\psfrag{m2q01}{}
%\centering
\includegraphics[width=0.45\textwidth,height=5cm]{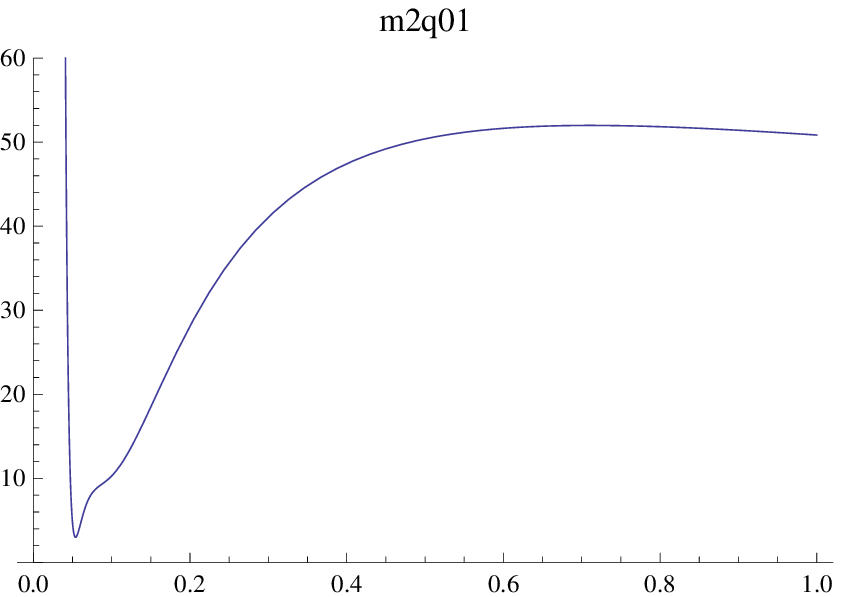}}
\subfigure[$M=10, Q=1$]{
\psfrag{M10Q1}{}
\includegraphics[width=0.45\textwidth,height=5cm]{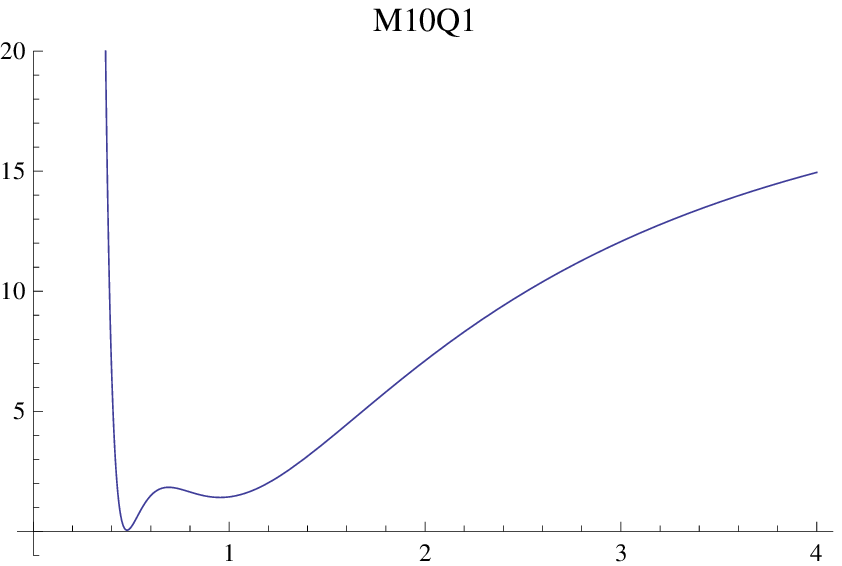}}
\caption{$V(r)$ for $N=\frac{6}{7}$}
\end{figure}

%%%%%%%%%%%%%%%%%%%%%%%%%%%%%%%%%%%%%%%%%%%%%%%%%%%%%%%

\section{The behaviour of the Radial Equation at Asymptotia and near Horizon}\label{Radial}
We analyze the radial equation~\eqref{second order} in two distinct radial regions, \textit{viz.}, near the horizon $(r^* \rightarrow -\infty)$ and at asymptotia $(r^* \rightarrow +\infty)$. In the asymptotic region $(r^* \rightarrow +\infty)$, Eq.~\eqref{second order} can written approximately as,
\begin{equation}\label{asymptotic}
\frac{\ud^2 R_\infty}{\ud{r^*}^2} + \omega^2 R_\infty = 0.
\end{equation}
This equation can be solve trivially as
\begin{equation}\label{r_infty}
 R_{\infty} = \ue^{i\omega r^*} + \mathcal{R} \ue^{-i \omega r^*}
\end{equation}
The first term represents the ingoing wave and the second term corresponds to the reflected wave. Thus $\mathcal{R}$ plays the role of the reflection coefficient. 
Near the horizon $\left(r^* \rightarrow -\infty\right)$, Eq.~\eqref{second order} can be written approximately as,
\begin{equation}\label{horizon}
\frac{\ud^2 R_H}{\ud{r^*}^2} + \omega^2 R_H = 0.
\end{equation}
This equation again can be solved trivially as
\begin{equation}\label{r_horizon}
 R_H=\mathcal{T} \ue^{i\omega r^*},
\end{equation}
where we have imposed the boundary condition that of the two solutions of the equation, only the ingoing wave is the physical one. Thus  $\mathcal{T}$ is termed as the transmission coefficient. 

If we now calculate the Wronskian for both the cases, we obtain
\begin{equation}
 \begin{array}{l}
 \mathcal{W}(+\infty) = -2i\omega(1-|\mathcal{R}|^2)\\
 \mathcal{W}(-\infty) = -2i\omega |\mathcal{T}|^2.
 \end{array}
\end{equation}

The equality of the two Wronskians immediately gives the following relation between the two coefficients --
\begin{equation}
 |\mathcal{T}|^2 = 1- |\mathcal{R}|^2.
\end{equation}

This shows that the transmission and the reflection coefficients are both less than unity -- irrespective of the presence of the dilatonic or the electric charge of the black hole. Thus superradiance mode does not exist for the black hole under consideration. It is worthwhile to spend a few lines about the phenomenon of superradiance at this point. This phenomenon is the wave analogue of energy extraction from black hole via Penrose process. Energy extraction via this process is only possible if and only if the black hole is rotating \textit{i.e.}, ergosphere of the black hole exists. The reason is that while one goes from outside to inside of the ergosphere, the timelike Killing vector becomes spacelike~\cite{wald}. The wave extracts energy from the `spinning-down' of the rotating black hole. This phenomenon was first discovered by  Zeldovich~\cite{zeldovich}. Later it was investigated by Starobinsky~\cite{starobinski} and Misner who gave the name `Superradiance'~\cite{dewitt}. The quantum version of superradiance can be likened to be the `stimulated emission of radiation' from the black hole whereas the Hawking radiation is the `spontaneous radiation'~\cite{basakmajumdar}. The black hole under consideration is non-rotating -- \textit{i.e.}, it has no ergo-region. Thus no superradiance is expected. Our calculation corroborates this.

However, Shiraishi~\cite{shiraishi} and Ali~\cite{ali} claimed that superradiance is possible if the black hole is ``a charged and/or rotating'' one. Shiraishi analyzed charged dilatonic black hole. Earlier Raychaudhuri et.al.~\cite{raychaudhuri} showed that the field of the dyadosphere of a black hole does not  superradiate even if it is charged -- as opposed to Shiraishi's claim. It is also clear from Chandrasekhar's analysis of Reissner-Nordstr\"om black hole that for this class of black holes superradiance is not possible. All this is supported by our calculation. Any further comment on Shiraishi's claim is beyond the scope of this paper and we reserve this for further communication.

%%%%%%%%%%%%%%%%%%%%%%%%%%%%%%%%%%%%%%%%%%%%%%%%%%%%%%%%%%%%%%%%%%%%%

\section{Summary}
In this paper we have investigated  the scalar perturbation of static charged dilatonic black solution in $(2+1)$-dimensions proposed by Chan and Mann. In doing so, we have written down the Klein-Gordon field equation representing the scalar wave in the curved spacetime of Chan-Mann black hole background metric. The angular and the radial parts  of the resulting equation are separated assuming a particular ansatz. Using tortoise coordinate, the radial equation is written in form of a one-dimensional Schr\"odinger type equation. The relation between the transmission and reflection coefficients for the scalar wave is established to show the absence of superradiance contrary to claims of some authors in similar type of situation.

%%%%%%%%%%%%%%%%%%%%%%%%%%%%%%%%%%%%%%%%%%%%%%%%%%%%%%%%%%%%%%%%%%%%%

\section*{Acknowledgment}

BR likes to thank Inter-University Center for Astronomy and Astrophysics, Pune, India for their hospitality and facilities extended to him during his visits under Associateship programme.

%%%%%%%%%%%%%%%%%%%%%%%%%%%
%%%%Bibliography%%%%

%%%%%%%%%%%%%%%%%%%%%%%%%%%%%%%%%%%%%%%%%%%%%%%%%%%%%%%%%%%%%%%%%%%%%%%%%%%%%%%%%%

\end{document}